\documentstyle[preprint,tighten,eqsecnum,epsf,aps]{revtex}

\def\beq{\begin{equation}}   \def\eeq{\end{equation}}
 
\newcommand{\ra}{\rightarrow}

\begin{document}
\draft
\title{How quark hadron duality in QCD may work.}
\author{B. Blok\thanks{ Talk given at  QCD-97 Euroconference by B.Blok,
based on joint work by B.Blok, M. Shifman and D.-X. Zhang,
preprint UMN-TH-1542/97; TPI-MINN-13/97-T}} 
\address{Department of Physics, Technion -- Israel Institute of 
Technology, Haifa 32000, Israel}

\maketitle

\begin{abstract}
We pursue  the issue of the local quark-hadron duality 
at high energies in two-  and four-dimensional
QCD. A mechanism of the dynamical realization of the 
quark-hadron duality in two-dimensional QCD in the limit of  large 
number of 
colors, $N_c\rightarrow\infty$, (the ' t Hooft model) is investigated. 
We argue that a similar mechanism of 
dynamical smearing may be relevant in 
four-dimensional QCD. Although particular details of our results are 
model-dependent (especially in the latter case), the general features
 of the duality implementation conjectured previously get further 
support.
\end{abstract}
\section{Introduction}

 In the recent years the focus of applications of the operator product
expansion (OPE) \cite{Wils1} has shifted towards the processes with 
the 
essentailly Minkowskean kinematics. Perhaps, the most well-known 
example
is the theory of the inclusive decays of heavy flavors (for a review 
see e.g.
Ref. \cite{Shif1}). This fact, as well as the increasingly higher 
requirements
to the accuracy of predictions, puts forward the study of the quark$-$
hadron 
duality as an urgent task.

A detailed definition of the procedure which goes under the 
name
of the quark$-$hadron duality (a key element of every calculation 
referring to 
Minkowskean quantities) was given in Refs. \cite{Shif1,Chib1}. In a 
nut shell,
a {\em truncated} OPE is analytically continued, term by term, from 
the 
Euclidean to the Minkowski domain \footnote{Moreover, usually one 
deals 
with the practical version of OPE, see Ref. \cite{Chib1} for further 
details}. A 
smooth quark curve obtained in this way is
postulated to coincide at high energies (energy releases) with the 
actual
hadronic cross section. 

If duality is formulated in this way, it is perfectly obvious that
deviations from duality must exist. In Ref. \cite{Shif1} it was shown
that one of the sources is the asymptotic divergence of high orders in 
the 
power series. If we knew the leading asymptotic behavior of the high 
order
terms in the power series we could predict the pattern of the 
duality-violating 
contributions at high energies. Unfortunately, very little is known 
about this
aspect of OPE, and we have to approach the problem from the other 
side --
either by modelling the phenomenon \cite{Chib1} or by studying 
some
general features of the appropriate spectral densities. One can also 
try to 
approach the problem purely phenomenologically (for recent 
attempts
see e.g. \cite{Blok1,Alta1,Blok2}).

An illustrative spectral density, quite instructive in the studies of the
issue of the quark-hadron duality,  was suggested in \cite{Shif1},
\beq
{\rm Im}\Pi = \mbox{Const.} N_c \, \pi\, \sum_{n=1}^\infty \delta 
({\cal E} - n)
\label{modim}
\eeq
where
$$
{\cal E} = \frac{s}{\Lambda^2}\, ,
$$ 
and  we  droped an inessential constant in front of the 
sum.
The color factor $N_c$ is singled out for convenience.
The imaginary part above represents, for positive values of $s$, a 
sum of 
infinitely narrow
equidistant resonances, with equal residues. It defines $\Pi (q^2)$ 
everywhere
in the complex plane $q^2$, through the standard dispersion relation, 
up to an 
additive constant which can be adjusted arbitrarily. It is not difficult 
to see
that the corresponding correlation function
\beq
\Pi (q^2) = -N_c \left[ \psi (\varepsilon ) + 
\frac{1}{\varepsilon}\right]
\eeq
where $\psi$ is the logarithmic derivative of Euler's gamma function, 
and
$$
\varepsilon = -\frac{q^2}{\Lambda^2} = - {\cal E} \, .
$$
In the Minkowski domain ${\cal E}$ is positive, in the Euclidean 
domain
$\varepsilon$ is positive. Then, the asymptotic expansion of $\Pi 
(q^2)$
in deep Euclidean domain is well known,
\beq
\Pi (q^2) \ra -N_c\left[ 
\ln \varepsilon +\frac{1}{2\varepsilon} - \sum_{n=1} 
\frac{B_{2n}\varepsilon^{-2n}}{2n}\right]
\eeq
where $B_{2n}$ are the Bernoulli numbers. At large $n$ they  grow 
factorially, 
as
$B_{2n}\sim (2n)!$ (see \cite{Wang}, page 23), and are sign 
alternating.

Although the spectral density (\ref{modim}) is admittedly a model, it 
was 
argued \cite{Shif1} that a similar factorial growth of the coefficients 
in the 
power (condensate) series is a general feature.

The spectral density (\ref{modim}) may be relevant in the limit
of the large number of colors, $N_c\ra\infty$, when all mesons are 
infinitely
narrow. This limit is not realistic, however, and, moreover, in this 
limit
the {\em local} quark-hadron duality, as we defined it, {\em never} 
takes place
since even at high energies the hadronic spectral density never 
becomes 
smooth. One can smear it by hand, of course, but then deviations 
from the local duality
will be determined not only by the intrinsic hadronic dynamics, as is 
the case 
in the real world, but also by details of the smearing procedure
-- the weight function chosen for smearing, the interval of smearing 
and so on. 
In the actual world the smearing occurs {\em dynamically}, since at 
high 
energies
the resonance widths become non-negligible. The limits of 
$E\ra\infty$
and $N_c\ra\infty$ are not interchangeable. 

In this work we study more realistic (dynamically smeared) spectral 
densities
compatible with all general properties of Quantum Chromodynamics.
Starting from  infinitely narrow resonances, as in Eq. (\ref{modim}),  
we then 
introduce finite widths, ensuring smooth behavior. Various 
dynamical regimes 
leading to specific patterns of duality violations are considered,
with a special emphasis on the high-order asymptotics of the power 
series.
Technically, in the first part of the paper the problem of duality is 
analyzed in
the two-dimensional 't Hooft model \cite{tHoo1} (see also
\cite{CCG,E1,E2}).
We see that there indeed exists the 
dynamic smearing of the spectral density.
This smearing occurs already if one takes into account the fact that
 the resonance widths are nonzero and calculates the relevant spectral 
densities in the Breit-Wigner approximation. We see that there are  
two characteristic scales in the problem. Starting from the first scale 
 the spectral density is well approximated by the oscillations round
the average value given by the OPE asymptotics. Starting from the second
scale the oscillations become small and local duality is valid.

The $\psi$-function model was originally suggested in Ref.
\cite{Shif1} for the heavy-light quark systems. In Ref. \cite{Zhit1} it 
was 
noted that it was more appropriate for the light-light quark systems
since in the heavy-light  systems the resonances are not expected
to be equidistant. Straitforward quasiclassical estimates
yield in this case that the meson energies (measured from the heavy 
quark mass) asymptotically scale as $\sqrt{n}$. (More elaborated 
models, e.g. the one of Ref. \cite{CDN}, based on a relativistic linear 
potential, yield not only the $\sqrt{n}$
law, but also the coefficient in front of $\sqrt{n}$.) 
In the present paper we further develop the $\psi$-function model
adapting it for the light-quark systems. 

\section{resonance widths in t'Hooft model}

 Our main original contribution to t'Hooft model is the  calculation of
 the first 
nonvanishing corrections to the 
widths 
of the resonances of t'Hooft model. Then, using these widths, we 
shall 
find the asymptotic behavior of  the polarization operator of two 
scalar currents. This resonance-saturated polarization operator
will be referred to as phenomenological. We will confront it with
the truncated power expansion. The difference between these two 
expressions presents duality violation.

It was shown by t'Hooft that the 2d QCD 
 is exactly solvable in the 
limit
$N_c\rightarrow\infty$. The bound state spectrum includes an 
infinite 
number of bound states whose masses $m_n$ lie on the almost linear 
trajectory.
The properties of these bound states are described by the t'Hooft 
equation 
\begin{equation}
\mu^2_n\phi_n(x)=\frac{(\gamma^2 -1)}{x(x-1)}-
\not\!\int^1_0\frac{\phi_n(y)dy}{(x-
y)^2}.
\label{equation}
\end{equation}
Here the integral is understood as a principal value,
The variable $x$ is the momentum fraction of the meson carried by
the antiquark (in the infinite momentum frame), while $1-x$ is that
of the quark; 
$\phi_n(x)$ is the wave function of the $n$-th bound state,
$\mu^2_n=m^2_n/\mu^2$
\par The integral equation \ref{equation} must be solved with the boundary
conditions:
\beq
\phi_n (x)\rightarrow x^\beta (1-x)^\beta 
\label{boundary}
\eeq
for $x\rightarrow 0,1$. Here  
\beq
\pi\beta{\rm ctg} (\pi\beta ) =1-\gamma^2
\label{b2}
\eeq
and $\gamma^2=m_q^2/\mu^2$, $m_q$ is the mass of the quark.
\par Below we shall be interested in the massless case $\gamma =0$ and 
in the number of flavors equal to 1.
In this case the asymptotic behavior of
$\mu_n^2$ {\em versus} $n$ is given by equation 
\begin{equation}
\mu^2_n= \pi^2 n \left( 1+O(\ln (n)/n + ... \right)\, .
\label{masses}
\end{equation}

 As for the wave functions $\phi_n (x)$ , their calculation is quite 
complicated
even numerically,
especially in the massless case.
 The recent calculations 
\cite{KS}, exploiting a new improved numerical procedure (spline 
method), 
show that for the massless case these wave functions are very 
accurately 
approximated by 
\begin{equation}
\phi_n(x)=\sqrt{2}{\rm cos} (\pi n x)\, .
\label{wf1} 
\end{equation}
%The wave  functions (\ref{wf1}) differ from the wave functions used 
%by t'Hooft in his classic paper 
%\begin{equation}
%\phi_n^t(x)=\sqrt{2}{\rm sin}(\pi n x)
%\label{wf2}
%\end{equation}
% for arbitrary quark masses.
Note that the wave functions \ref{wf1} satisfy the proper boundary conditions 
for 
the 
massless case.
% Indeed, for nonzero masses, $\beta\ne 0$, and evidently 
%the right boundary conditions are
%\beq
% \phi_n (0)=\phi_n (1)=0
%\label{bq2}
%\eeq
%On the other hand for the massless case $\beta$ is zero.
% From eq. \ref{b2} it follows in the limit 
% $\beta\rightarrow 0$ that the boundary conditions \ref{boundary} become
%\beq
%\phi_n (0)=1, \phi_n (1)=P
%\label{bq3}
%\eeq
%Here $P=1$ for the states with even parity and $P=-1$ for the states with 
%odd parity.
%These are just boundary conditions satisfied by the wave functions
%\ref{wf1} .
\par Let us now calculate the meson widths.
 As it was already mentioned, in the limit 
$N_c\rightarrow 
\infty$ the bound states in the t'Hooft model are stable, their 
widths vanish. However, once one takes into account the leading
$1/N_c$ correction, the resonances begin to decay. In the first 
order in $1/N_c$ expansion there are only two-particle decays
$a\rightarrow b+c$. The relevant coupling constatnts $g_{abc}$ are 
given by
\cite{CCG,B,KS,E1,E2} :
\begin{equation}
g_{abc}=\frac{(g^2N_c)}{\pi}\sqrt{\pi/N_c}
(1-(-1)^{(\sigma_a+ \sigma_b+ \sigma_c)})(f^+_{abc}+f^-_{abc})
\label{decayconstant}
\end{equation}
Here $\sigma_a$ is the parity of the $a$-th resonance.
The constants $f^{\pm}_{abc}$ are determined from the following 
expressions:
$$
f^{\pm}_{abc}=\displaystyle\frac{1}{1-
\omega_{\pm}}\int^{\omega_\pm}_0
\phi_a(x)\phi_b(x/
\omega_\pm)\Phi_c\left( \frac{x-\omega_\pm}{1-\omega_\pm}
\right)dx \, ,
$$
\begin{equation}
-\frac{1}{\omega_\pm}\int^1_{\omega_\pm}\phi_a(x)\Phi_b(x/
\omega_\pm)
\phi_c\left( \frac{x-\omega_\pm}{1-\omega_\pm}\right)dx.
\, .
\label{c}
\end{equation}
Here $\omega_{\pm}$ are  two roots of the algebraic equation 
corresponding to the mass-shell condition,
\begin{equation}
m^2_a=\frac{m^2_b}{\omega} +\frac{m^2_c}{(1-\omega )} \, .
\label{massshell}
\end{equation}
Two different idices of $\omega$ correspond to two different 
modes
of the two-dimensional decay: in the rest frame of the resonance $a$
the resonance $b$ can go to the right and the resonance $c$ to the 
left,  and {\em vice 
versa}.
\par The function $\Phi_a (x)$ is defined as
$$\Phi_a (x)=\int^1_0 dy \phi_a (y)/(x-y)^2.$$
It is easy to carry the calculation of the couplings $f_{abc}$ analytically
(see Ref. \cite{BSZ}) and express them via integral sinus and integral 
cosinus functions.
 Using the above expresions for the decay  couplings one can readily 
calculate the resonance 
widths 
in the leading $1/N_c$ approximation. They are given 
by
\begin{equation}
\Gamma_a=\frac{1}{8m^3_a}\sum_b\sum_c\frac{(m^2_a+m^2_b-
m^2_c)g^2_{abc}
}{\sqrt{I(m^2_a,m^2_b,m^2_c)}}\, ,
\label{sum}
\end{equation}
where $I$ is the standard ``triangular" function,
\begin{equation}
I(x,y,z)=\frac{1}{4}[m_a^2-(m_b+m_c)^2][m_a^2-(m_b-m_c)^2]\, .
\label{triangular}
\end{equation}
The sum in Eq. (\ref{sum}) runs over all mesons $b$ and $c$ whose 
masses are lighter than that of $a$. 

Details of our calculation of the resonance widths are 
described 
in  Ref. \cite{BSZ}. The calculation was done numerically  using the trial 
wave 
functions 
(\ref{wf1}). Our task was to establish the asymptotic behavior of the 
widths, as a function of the excitation number, at large values of
$n$, in the leading $1/N_c$ approximation (all widths are
proportional to $1/N_c$). After computing the overlap integral we 
performed the summation over $b$ and $c$. The result for the 
widths exhibits a remarkable pattern. The widths of the individual 
levels oscillate near a smooth square-root curve, see Fig. 1. This plot 
shows the width of the $a$-th state {\em versus} $a$, up to $a= 500$.
 The result of averaging over the 
interval 
of 20 resonances is depicted in Fig. 2. We see that the curve of the
resonance 
widths $\Gamma (a)\equiv \Gamma_a$ is very
well approximated by the function 
 \begin{equation}
\Gamma (a)=\frac{A\mu}{\pi^3 N_c}\sqrt{a}\left( 1+ {\cal O} 
(1/a)\right) \, ,
\label{widths}           
\end{equation}
where  $\mu^2=g^2N_c/\pi$. 
% More exactly, by fitting the whole 
%interval of $a$ up to $a=500$ we find that the best fit for the power 
%of $a$ is 
%$0.53\pm 
%0.03$. 

Since the square-root law (\ref{widths}) for the (averaged) widths is 
valid in such a large interval of the excitation numbers and 
turns out to be so accurate,  it seems to be doubtless that
this formula could be obtained analytically. This is an interesting 
question by itself, especially in the four-dimensional QCD.
Unfortunately, we were unable to find analytic solution so far.

 The numerical value of the constant $A$   is
% \footnote{If one  
%calculates the widths with  the wave functions (\ref{wf2}), the 
%square-root behavior of  
%Eq. (\ref{widths}) is intact,  but with the value of the  constant $A$ is
%different,  $A'\sim 0.007$, i.e.
%$\sim 50$ times 
%smaller. This fact indicates that the square-root law is not sensitive 
%to the 
%precise form of the wave functions, while the  value of the 
%coefficient $A$ is.
%}
\begin{equation}
A\sim 0.44. 
\label{constant}
 \end{equation}
Below the above result for the (averaged) widths will be used for
determining of the asymptotic behavior of the polarization operator. 

Concluding this section let us note that  the same square-root
was reported previously 
 in Ref. \cite{B}. We failed to reproduce the arguments of this work 
leading to the square-root law, however. Moreover, what is even 
more important, the constant $A$ in Ref. \cite{B} is 
claimed to be proportional to $1/\sqrt{m}$ (!), and, 
thus, 
blows up for  massless quarks. This poses perplexing questions.
The coincidence looks completely accidental.

\section{Quark-hadron duality in 2d QCD}
Once we had found the resonance widths, we can calculate the 
polarisation operator in the Breit-Wigner approximation. Here we 
shall 
consider the most interesting case of the polarisation operator of the 
two 
scalar currents. Let us start from this polarisation operator in the 
$N_c\rightarrow\infty$ limit \cite{CCG,Z}.
\par The polarisation operator is given by the 
correlator
 \begin{equation}
\Pi (q^2)=i \int d^2x {\rm e}^{iqx}<0\vert T\{j(x),j(0)\}\vert 0>
\label{Pol}
\end{equation}
Here j(x) is the scalar current:
$$j(x)=\bar q(x) q(x)$$
and q(x) is the quark field.
It was shown in Ref. \cite{CCG} that this polarisation operator can be 
calculated explicitly in the $N_c\rightarrow\infty$ limit 
and is given by 
\begin{equation}
\Pi(q^2)=\frac{N_c}{\pi}
\sum^{n=\infty}_{n=1}\frac{f^2_n}{q^2-m^2_n+i\epsilon}
\label{pol1}
\end{equation}
It was shown in Ref. \cite{Z} that in order for eqs. \ref{pol1}
and the perturbation theory for QCD$_2$
to be compatible, the coeffficients $f_n$ for sufficiently large n must 
be 
independent of n. Then, taking into account the linear dependence 
of the mass squared on n one can approximate the polarisation operator 
\ref{Pol} for sufficiently large $q^2$ by $\psi$-function \cite{Z}:
\begin{equation}
\Pi (q^2)-\Pi(0)\sim \frac{N_c}{\pi}
\psi(-q^2/\pi^2\mu^2 ) 
\label{psi}
\end{equation}
Here 
$\lim_{n\rightarrow\infty} f^2_n\sim f^2=\mu^2\pi^2$. 
\par Consider now what will happen if we take into account the 
finite 
widths of the resonances. We shall calculate the polarisation operator 
in the Breit-Wigner approximation. In order to do it we  
shall
first find the inverse propagator of the n-th bound state:
\begin{equation}
\Pi_n(q^2+i\epsilon )=q^2-m^2_n+\Sigma (q^2+i\epsilon )
\label{sigma}
\end{equation}
Using the procedure described in Ref. \cite{B} it is easy to prove 
(see Ref. \cite{BSZ} for details) that
%{\rm Im}\Sigma(q^2+i\epsilon)=\sqrt{q^2}\Gamma (q^2)
%\label{im}
%\end{equation}
%Here $\Gamma (q^2)$ is the resonance width in terms of the $q^2$ 
%variable:

%$\Gamma (q^2)=\frac{B}{N_c} \sqrt{q^2}$
%where $B=\frac{A}{\pi^4}$
%The latter equation is valid starting from the some cut-off value :
%$q^2\ge \bar q^2$.
%One then obtains \cite{B} 
\begin{equation}
\Sigma (q^2+i\epsilon)\sim +\frac{Bq^2}{N_c\pi}{\rm Log}( 
-\frac{q^2}{\bar q
^2}-i\epsilon ) 
\label{in}
\end{equation}
The latter equation is valid for $q^2\ge \bar q^2$,
where  $\bar q^2$ is some intermediate scale, and 
$B=A/\pi^4$.
\par Then one obtains for the polarisation operator
\begin{equation}
\Pi (q^2)=\frac{N_c}{\pi}
\sum_n \frac{f^2_n}{q^2-m^2_n+\frac{Bq^2}{N_c\pi}{\rm Log}(-
\frac{q^2}{\bar q^2}-i\epsilon)}
\label{pol3}
\end{equation}
(for the nonphysical sheet).
Now we can use the known aymptotic behaviour of $m^2_n$ and 
$f^2_n$ to 
sum the series \ref{pol3}.
We immediately obtain:
\begin{equation}
\Pi (q^2)-\Pi (0)\sim \frac{N_c}{\pi}
\psi (-\frac{q^2+\frac{Bq^2}{N_c\pi}
{\rm Log} (q^2/\bar q^2)+iBq^2/N_c}{\mu^2\pi^2 })
\label{pol4}
\end{equation}
In other words we obtain that the polarisation operator on the 
unphysical
sheet is the psi-function of the complex argument. This is our 
main result.
%\par Note that our expressions for the polarisation operator are 
%consistent 
%with the asymptotics at $q^2\rightarrow\infty$ given by 
%perturbation 
%theory:
%\begin{equation}
%\lim_{q^2\rightarrow \infty} \Pi (q^2)=\frac{g^2N_c}{\pi^3\lambda^2}
%\log q^2 +  ....
%\label{as}
%\end{equation}
%The imaginary part tends to constant:
%\begin{equation}
% \lim_{q^2\rightarrow\infty}{\rm Im}\Pi 
%(q^2)=g^2N_c/(\pi^3\lambda^2)+ ...
%\label{eqs}
%\end{equation}
%These is just the leading behaviour of the perturbation theory.
In the limit $N_c\rightarrow\infty$ we clearly recover  eq. 
\ref{psi} .
\par We can  now study the duality violation and the 
operator
product expansion for the polarisation operator \ref{Pol} .
\par Let us start from the imaginary part. The are two possible 
patterns of behaviour of the imaginary part of the polarisation 
operator 
as a function of $q^2$: Euclidean kynematics corresponds to $q^2\le 
0$ 
and Minkowskean kinematics corresponds to  $q^2\ge 0$. Note that in 
the 
Eclidean domain the imaginary part of the polarisation operator
\ref{pol3} is zero. Indeed, the logarithm in the argument of the $\psi$
function  does not have any imaginary part, and the entire 
polarisation operator is real. Thus, the imaginary part of the 
polarisation operator is zero for $q^2\le 0$ as it must be.
Consider now the imaginary part in the Minkowskean domain
(i.e. the spectral density). 
In the limit $N_c\rightarrow\infty$ the imaginary part of the 
polarisation 
operator is  the sum of the delta-functions, corresponding to the 
poles 
of the psi-function:
\beq
{\rm Im} (\Pi (q^2)-\Pi (0))\sim\frac{f^2N_c}{\pi}\sum_n\delta (q^2-n
\mu^2 \pi^2).
\label{einsof}
\end{equation}
However, once we take into account the leading $1/N_c$ correction, 
the 
situation immediately changes. Instead of the sum of the delta 
functions,
one obtains the oscillations of the imaginary part round the constant
value $N_c/\pi$ with decreasing  amplitude.
Note that the latter is just the 
average value of the imaginary part of \ref{einsof} over the interval 
$q^2$. Thus we see the example of dynamic smearing: instead of 
infinite 
peaks, QCD dynamically produces decreasing oscillations round average
smooth function.
\par Let us now check that this is indeed the case for the concrete 
example of our polarisation operator \ref{pol3} . In order to find the
relevant  
imaginary part explicitly, we shall make use of the reflection 
property of $\psi$-function:
\beq
\psi (z)= \psi (-z)-\pi{\rm ctg} (\pi z)-1/z.
\label{ref}
\eeq
Then, using eq. \ref{pol3}
one obtains
\begin{eqnarray}
&{\rm Im} (\Pi (q^2)-\Pi (0))=
\displaystyle\frac{N_c}{\pi}\{({\rm Im}\psi ((q^2+
\frac{Bq^2}{N_c\pi}\log (q^2/\bar 
q^2)+\frac{iBq^2}{N_c})/(\pi^2\mu^2 ))\cr
&+\displaystyle{\rm Im}\pi
{\rm ctg}(\pi(q^2+\frac{Bq^2}{N_c\pi}\log {q^2/
\bar q^2 }+iBq^2/N_c)/(\pi^2\mu^2))\cr
&+\displaystyle{\rm Im}
 (\frac{\pi^2\mu^2}{q^2+\frac{Bq^2}{N_c\pi}\log{q^2/\bar q^2}+iBq^2/N_c}
)\}\cr
\label{refim}
\end{eqnarray}
\par  It is easy to see that a sum of the first and the third
 terms in eq. \ref{refim} is the 
smooth 
function of $q^2$ (for positive $q^2$). Numerically these functions 
are well approximated by the OPE (asymptotics).
\par Consider now the second term
in eq. \ref{refim}. For $N_c\rightarrow \infty$ this term 
has simple poles in the points where $$ q^2=\mu^2 \pi^2 n,$$
i.e. the imaginary part of this term is just the sum of the delta 
functions 
\ref{einsof} .
\par Consider now the case of the large but finite $N_c$. Then for 
$q^2\ge \pi^2\mu^2$ the
imaginary part of the 
 polarisation operator \ref{refim} is well 
approximated by the    
 imaginary part of the ctg term in eq. 
\ref{refim}:   
\beq
{\rm Im} (\Pi (q^2)-\Pi (0))\sim -N_c\frac{  {\rm sh} (2y)}{{\rm ch} (2y)-
{\rm cos} (2x)}).
\label{trig}
\end{equation}
This means that for $y\rightarrow\infty$
\beq
{\rm Im}{\rm \Pi} (q^2) 
 \sim
-N_c(1+2\exp{(-2y)}{\rm cos} (2x) +...
\label{exp}
\eeq 
 Here $x=q^2/(\pi\mu ^2)+O(1/N_c)$
and $y=Bq^2/(N_c\pi\mu^2)$.
\par  We can now see the behaviour of the imaginary part of the 
polarisation operator.
For $q^2\ge \pi^2\mu^2$
 this behaviour is well approximated by 
the ctg term  (eq. \ref{trig}).
 The poles in the imaginary part
of eq. \ref{trig} disappear for 
arbitrary nonzero y, i.e. for every finite $N_c$. Instead we have 
peaks 
with the amplitude given for $q^2\sim \mu^2\pi^2$ by
$N_c/(Bq^2)$
The amplitude of these peaks goes to $\infty$ as $N_c\rightarrow 
\infty$.
\par  For
\beq
q^2\ge N_c\pi\mu^2/(2B) 
\label{r}
\eeq
the asymptotic behaviour will be given by eq. \ref{exp}.
 The oscillations
decrease exponentially
 and local duality is established.
At these $q^2$ one may use the OPE. For  $q^2$
smaller than the scale given by eq. \ref{r} the oscillations dominate
and we can speak only about global duality.  
For $q^2$ much larger than this scale one can safely use the OPE.
The corresponding inaccuracy decreases exponentially as 
${\rm exp}(-2q^2B/(N_c\pi\mu^2))$.
\par Numerically the typical behaviour of the spectral density 
is depicted in Fig. 2 for $\mu^2 =0.01, N_c=10$.
The function F1 depicts the
(normalised) behaviour of the full $\psi$ function expression 
for the spectral density, F2-of the ctg term in eq. \ref{pol4} 
and F3-the exponentially decreasing term F3.
\par Consider now the full polarisation operator \ref{pol3}. For  
$q^2
\le 0$ the polarisation operator is evidently given by the $\psi $ 
function 
of the positive real argument and is a smooth function with no 
singularities. It is well approximated by its asymptotic expansion 3.12:
\begin{eqnarray}
&\Pi (q^2)-\Pi (0)=\displaystyle \frac{N_c}{\pi}(\log{(-
q^2/\mu^2\pi^2)}\cr
&\cr
&+\displaystyle \sum \frac{(\mu\pi)^{4n-4}}{2n}\frac{ 
B_{2n}}{q^{2n}
}+O(1/N_c))\cr
\label{42}
\end{eqnarray}
Here $B_{2n}$ are the Bernulli numbers.
Note that the expansion \ref{42} is not Borel summable. Indeed, for 
large n
the ratio $a_{n+1}/a_n$, where $a_n$ is the n-th term of the series 
\ref{42}, becomes $\sim 1$ staring from $n\sim -q^2/(\mu^2\pi 
)$ This means
that the asymptotic expansion has the intrinsic uncertainty
\beq
\Delta \Pi (q^2)\sim \exp{(\frac{2q^2}{\mu^2\pi} 
\log{-2q^2/\lambda^2}) }.
\label{43}
\eeq
Note that it is senseless to continue analytically the exponentially 
decreasing term \ref{43} into the Minkowski domain, although it has 
the 
multiplier that begins to oscillate upon such continuation.
This teaches us that it is extremely dangerous to continue different 
parts 
of the polarisation operator analytically from Euclidean to 
Minkowski
domain.
\par The $1/N_c$ terms in the Euclidean domain are only small corrections
to the $N_c\rightarrow \infty$ expression and can be neglected numerically.
\par We can now carry the staightforward analysis of the large $q^2$ 
behaviour of the asymptotic expansion of the polarisation operator.
This is easily done in the same way as it was done for the imaginary 
part.

 We immediately see that in the Minkowski domain
the polarisation operator is given by the sum of the asymptotic 
expansion 
\ref{42}
(continued analytically into the Minkowski domain)
 and the oscillating part that for $q^2\ge N_c\pi\mu^2/(2B)$
is equal to
\beq 
\Pi^{\rm osc} (q^2)\sim 2\frac{N_c}{\pi}\sin{(2 
q^2/(\pi\mu^2))}
\exp{(-2Bq^2/(N_c\pi\mu^2 ))}.   
\label{oscil}
\end{equation}
We once again see an oscillating part with the exponentially 
decreasing 
amplitude of the oscillations that must be added to the smooth 
OPE function. 
\par We conclude that there are indeed two characteristic scales in 
the 
problem. For $q^2\ge \pi^2\mu^2$ the polarisation operator is 
dominated
by oscillations, and this oscillations begin to decrease exponentially 
at 
the scale given by eq. \ref{r}.  Starting from the latter scale
the amplitude of the oscillations become smaller than the leading 
term
in the OPE and there is sense to speak about local duality,
that establishes itself for $q^2$ much bigger than the scale of 
eq. \ref{r}. The corresponding inaccuracies decrease exponentially.
 On other hand the global duality holds for all values of 
$q^2$.  
\section{OPE and the polarisation operator in 4d QCD.}
 \par In this section we shall consider the OPE in 4d QCD. Once 
again, 
we consider the polarisation operator of 2 vector currents and 
calculate 
it for large absolute values of $q^2$ for both Euclidean and 
Minkowski 
domains.
\subsection{Polarisation operator of vector currents in the Veneziano 
model}
\par Our goal in theis subsection will be to study the polarisation 
operator of two vector currents:
\beq
\Pi_{\mu\nu} (q^2)
 =\int\exp{iqx}<0\vert T\{j_\mu (x) j_\nu (0)\}\vert0>d^4x.
\label{pol8}
\eeq
Here $j_\mu (x)$ is the vector current,
\beq
j_\mu (x)=\bar u\gamma_\mu d(x).
\label{current}
\eeq
Due to the conservation of vector current  the polarisation operator 
can be 
represented as 
\beq
\Pi_{\mu\nu} (q^2)= (q^\mu q^\nu -q^2g_{\mu\nu})\Pi (q^2).
\label{polarisation}
\eeq
The polarisation operator can be calculated in the Euclidean domain 
using 
the OPE:
\beq
\Pi (q^2)\sim \log{-q^2}+a_0/q^4+....
\label{ope}
\eeq
Usually in order to calculate $\Pi (q^2)$ in the physical, Minkowski 
domain 
we just analytically continue eq. \ref{ope} into the Minkowski
domain.
In particular, the logarithm acquires the constant imaginary part.
Then we get 
\beq
\Pi (q^2)\sim \pi +O(1/q^2)...
\label{im3}
\eeq
\par This is not the whole story however. Indeed, we know from  
our 
experience with QCD$_2$ that in addition to the smooth constant there 
may 
be oscillations.
\par The question of oscillations is especially interesting since in QCD
the imaginary part of the polarisation operator of vector currents is 
directly related to the 
cross-section of the $e^+e^-$ annihilation into hadrons.
\subsection{Polarisation operator and resonance widths}
\par We have seen already  the connection between the resonance masses 
and widths 
asymptotics and the asymptotic of the polarisation operator in the 
Breit-Wigner approximation in 2d QCD. The same connection exists in 
4d QCD.
The problem is, however, that we do not know how to calculate the 
relevant 
resonance masses and widths in a model independent way, like it 
was
 done in QCD$_2$.
We need to know the masses and widths of the radial excitations 
with the 
quantum numbers of $\rho$-meson created by the vector current. In 
fact, we 
are interested only in the asymptotic behaviour of these widths and 
masses. We shall calculate masses of these excitations using the 
Veneziano 
model. The relevant radial excitations all have the same spin and lie 
on the
daugter trajectories to the $\rho$ meson trajectory (see Fig. 4):
\beq
m^2_n=n/\alpha'.
\label{rad}
\eeq
Here $\alpha '\sim 1 $ GeV$^{-2}$.
\par Although there exists a working model for the resonance 
masses, 
that is in agreement with the experimentally observed Regge 
behaviour,
virtually nothing is known about the dependence of the resonance 
widths on 
the
number n of the excitation. Based on our experience for 2d QCD it is 
natural to assume the power like behaviour:  
\beq
\Gamma_n\sim \frac{A}{N_c} (m^2_n)^{1/2+\gamma}.
\label{width2}
\eeq
For $N_c\rightarrow\infty$ the widths $\Gamma_n\rightarrow 0$.
The constant A is some unknown proportionality constant. The 
parameter $\gamma$ 
parametrises our ignorance of the 4d resonance widths. The only 
known 
theoretical estimate of $\gamma$ was made by Green and Veneziano
\cite{GV} on the 
basis of the string model, and their result is $\gamma=-1/2$. Here 
we 
shall  parametrise the widths by the apriory arbitrary parameter $\gamma$. 
 Using eq. \ref{width2} one immediately obtains the $q^2$ dependence 
of the self 
energy part of the inverse meson propagator:
\beq
{\rm Im}\Sigma (q^2+i\epsilon)=\frac{A}{N_c}q^{2(1+\gamma)}.
\label{IM}
\eeq
Then one can use the dispersion relations 
 to obtain $\Sigma (q^2+i\epsilon)$.
In order to do it we shall need to take the integral
$$I=\frac{Aq^4}{N_c\pi}\int^{\infty}_{s_0}\frac{s^{'\gamma -1}}{s'-
q^2}$$.
As usual we calculate this integral for negative $q^2$ and then 
analytically continue to positive $q^2$. Making substitution $s'/(-
q^2)=v/(1-
v)$ we obtain that 
\beq
I=\frac{A(-q^2)^{(1+\gamma)}}{N_c\pi}(B(\gamma,1-\gamma)-
\frac{1}{\gamma}(x/(1+x))^{\gamma}F(\gamma ,\gamma ,1+\gamma 
, x/(1+x)).
\label{int}
\eeq
Here $x=-s_0/q^2$, and $F(a,b,c,z)$ is the hypergeometric function. 
Then for large $q^2$ and noninteger $\gamma$ one obtains:  
\beq
\Sigma (q^2)\sim -\frac{A(-q)^{2(1+\gamma 
)}}{N_c\pi}\frac{\pi}{{\rm sin} (
\pi\gamma )}=q^{2(1+\gamma )}({\rm ctg} (\pi\gamma )+i)(A/ 
N_c).
\label{eq6}
\eeq
Note that for the $\gamma\rightarrow 0$ one obtains 
\beq
\Sigma (q^2)= \frac{A}{N_c\pi} q^2\log{(-q^2/s_0)}.
\label{gamma}
\eeq
Finally for integer negative $\gamma$ one needs the dispersion 
relation 
without substractions:
\beq
\Sigma (q^2)=\frac{A}{N_c\pi}q^2\frac{1}{(\alpha 
'q^2)^{\vert\gamma\vert}}
\log (-q^2/s_0).
\label{tsigma}
\eeq
Note that for each case $\Sigma (q^2)$ is real for $q^2\le 0$ and 
acquires 
the imaginary part for $q^2\ge 0$.
\par Now one can immediately obtain the polarisation operator:
\begin{eqnarray}
\Pi (q^2)=&\displaystyle\Sigma_n\frac{f^2_n}{q^2-n/\alpha '+\Sigma 
(q^2+i
\epsilon )}\cr
&=\displaystyle f^2\alpha '
\psi (-(q^2+\Sigma (q^2+i\epsilon))\alpha ')\cr
\label{62}
\end{eqnarray}
Here we used the fact that, analogously to $QCD_2$
in order for the sum over resonances to be consistent with the 
leading 
asymptotic behaviour, $f^2_n$ must be independent of n for large n.
\par For $q^2\le 0$ (Euclidean domain)
the polarisation operator is well approximated by  OPE:
\beq
\Pi (q^2)\sim f^2 \alpha'\log (-q^2\alpha ')+
+O(1/q^2, 1/N_c)...
\label{e1}
\eeq
The fact that we evidently get here the terms of the order $1/q^2$  
 reminds us that some of our results are model dependent.
It is evident, that by changing $n$ dependence, i.e. taking into account
that $f_n$ are constant only for large n, one can also change 
the behaviour of subleading terms in OPE. We do not write these terms 
explicitly due to their model dependence.
\par Consider now the $q^2\ge  0$ case -Minkowski domain.  
We can 
analytically continue eq. \ref{e1} into that domain. The 
polarisation operator  evidently acquires the imaginary part:
\beq
{\rm Im} \Pi (q^2)\sim (f^2\alpha')\pi +...
\label{im5}
\eeq
\par All the functions that we obtain by analytical continuation of 
the 
OPE are evidently smooth. In addition, in the Minkowski 
kynematics one 
gets  the oscillating part. For sufficiently large $q^2$ and 
nonpositive $\gamma$.  
\beq
{\rm Im} \Pi (q^2)\sim \pi\alpha'f^2{\rm Im}\,{\rm ctg}(\pi (q^2\alpha' +
iy)) 
\label{im7}
\eeq
Here 
$$y=Aq^2\alpha '/N_c$$
for $\gamma =0$ and 
\beq
y=A\vert q^2\vert ^{1+\gamma}/(N_c{\rm sin}(\pi\gamma ))
\label{68}
\eeq
for noninteger $\gamma$.
For integer negative $\gamma$ 
\beq
y=(A/N_c)q^2 \frac{1}{(\alpha'q^2)^{\vert \gamma\vert}}
\label{69}
\eeq
\par Note that for negative $\gamma\le -1$ one gets nondecreasing 
oscillations. This shows that in real QCD one must have the constraint
\beq
\gamma\ge -1.
\label{g}
\eeq
Moreover, we also see that $$\gamma\le 0$$.
Indeed, we have seen that if $\gamma $ is positive, than there 
appears
the term in $\Sigma$ that leads to the rise of the real part
of the polarisation operator quicker than $q^2$ (although this 
term is suppressed by $1/N_c$). This will lead to the curving 
upwards
of Regge trajectories that looks quite unlikely at present.
\par Numerically, the behaviour of the 
polarisation operator and its imaginary parts are depicted in Figs.3
for the sample value  $A/N_c$=0.3 and $\gamma =-1/2$
($\gamma =0$ behaviour is very similar to the one given by Fig.2).
 In these figures we
depict as $F0$ the imaginary part of the $\psi$ function,
as $F1$ the ctg approximation,
 and as $F2$ the exponentially decreasing asymptotic
oscillations. 
 Already starting from $q^2\ge 1/\alpha'$ the full polarisation 
operator is relatively well approximated by ctg. 
Starting from the scale 
\beq
q^2\ge (N_c/(A\alpha'))^{1/(1+\gamma )}
\label{scale}
\eeq
the full polarisation operator operator is represented by the 
OPE plus exponentially decreasing asymptotics. For scales 
much larger than the one given by eq. \ref{scale} the local 
duality is restored and deviations from it are exponential. 
\section{Conclusion}
\par We now can answer the basic questions asked in the 
introduction.
First, how the local quark- hadron duality works, and, second  
when one can use the OPE.
\par For the 2d QCD one can treat these two questions in a model 
independent way (the deviations from the Breit-Wigner approximation
can be estimated and are actually irrelevant for the problem at
hand). We found the polarisation operator of two scalar currents
and studied its duality properties.
\par We see that 
in the Minkowski domain
there are two characteristic scales in the problem.
The first scale is $\Lambda_1=\mu\pi$,
the second is given by $\Lambda^2_2=N_c\pi\mu^2/(2B)$  
 For the momentum 
$q^2$ between these two scales the oscillations dominate, and only 
global duality works. For the  $q^2$, larger than $\Lambda_2$ 
 the local duality works.
The deviations from local duality are only exponential.
 Consequently, for scales $q^2\ge\ge \Lambda^2_2$ one can use the local duality 
to calculate physical quantaties due to the exponential decrease of 
the 
amplitude of the oscillations.
 \par We thus can use the OPE to calculate global averages, i.e.
to calculate some integrals over the interval of $q^2$ much bigger 
than 
the distance between the resonances for almost all values of $q^2$.
However the use of OPE based on local duality is much more 
restricted, and we can say now that there is a dynamically generated 
scale in the problem starting from which the local duality is valid.
\par Moreover, we have seen that in the Minkowski domain 
there are always corrections to OPE, in the best case 
($q^2\ge\Lambda_2^2$ these corrections decrease exponentially.)
\par  The next question is what can be learned for 4d QCD.
Here we have seen that if Veneziano model and power like behaviour 
of the resonance widths are indeed true, the local duality 
is established for large $N_c$
in the same way as in 2d QCD. Once again, using this 
time Veneziano model, we see there two relevant scales
are $1/\alpha'$ and  $(N_c/(A\alpha'))^{1/(1+\gamma )}$ given by eq. 
\ref{scale}. Between these scales the spectral density 
is dominated by oscillations with decreasing amplitudes. For the 
$q^2$ larger than these scales the spectral density behaves 
like the OPE (analytically continued, term by term, as discussed in 
introduction) plus the exponentially decreasing oscillating term,
that decreases like $exp (-2A\alpha'q^{2(1+\gamma )})$
where A is some proportionality constant. 
\par  We can compare ourt results with the predictions of instanton 
vacuum model \cite{Chib1}. We see, similar to that model that there 
are three characteristic areas of momenta squared, divided by two scales.
The deviations from duality (i.e. corrections to the OPE)  decrease
exponentially for the scales larger than the biggest of these two scales,
like in Ref. \cite{Chib1}.

\newpage
\onecolumn

\begin{figure}%[htb]
\vspace{5cm}
{\epsfxsize5.0in\epsfbox{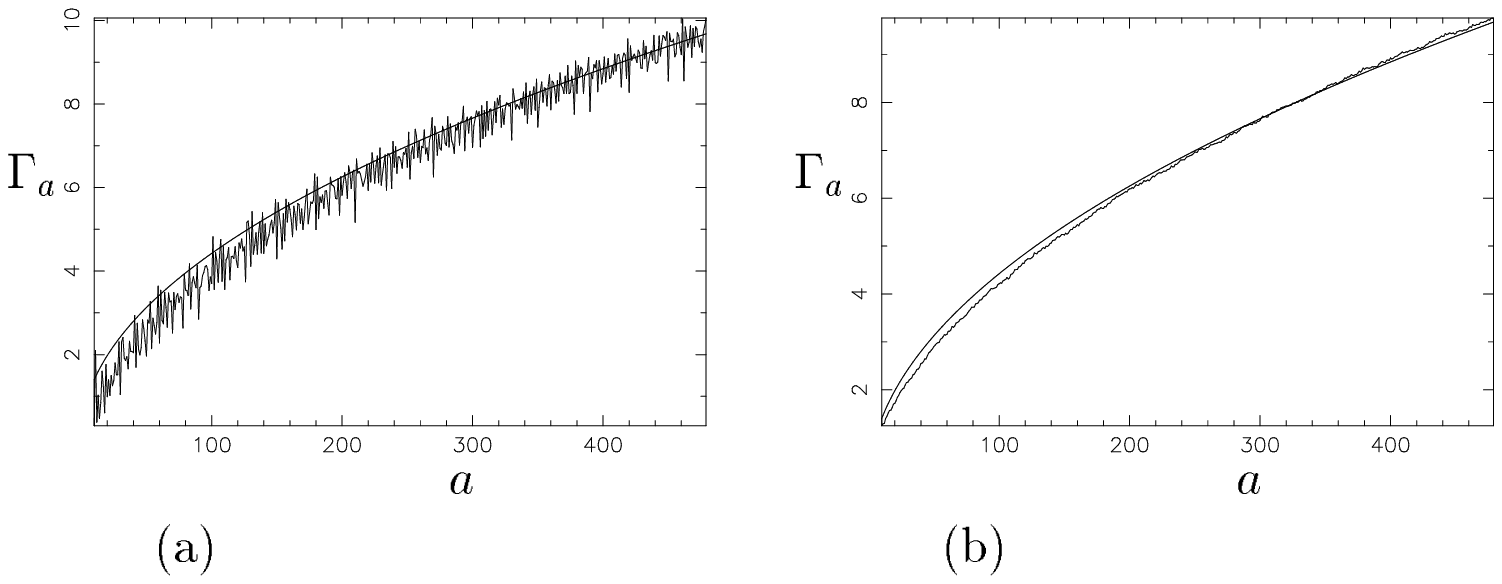}}
\caption{Comparision of the  decay width with the trial function.
The oscillating curves are widths $\Gamma_a$ 
of resonances {\it a}'s in (a),
and $\Gamma_a$ averaged over 20 resonances in (b),
while the smooth curve is $\Gamma_a=0.442a^{0.5}$.
(All widths are given in the unites $\mu/(\pi^3 N_c)$}
\end{figure}

\newpage

\begin{figure}[htb]
{\epsfxsize3.0in\epsfbox{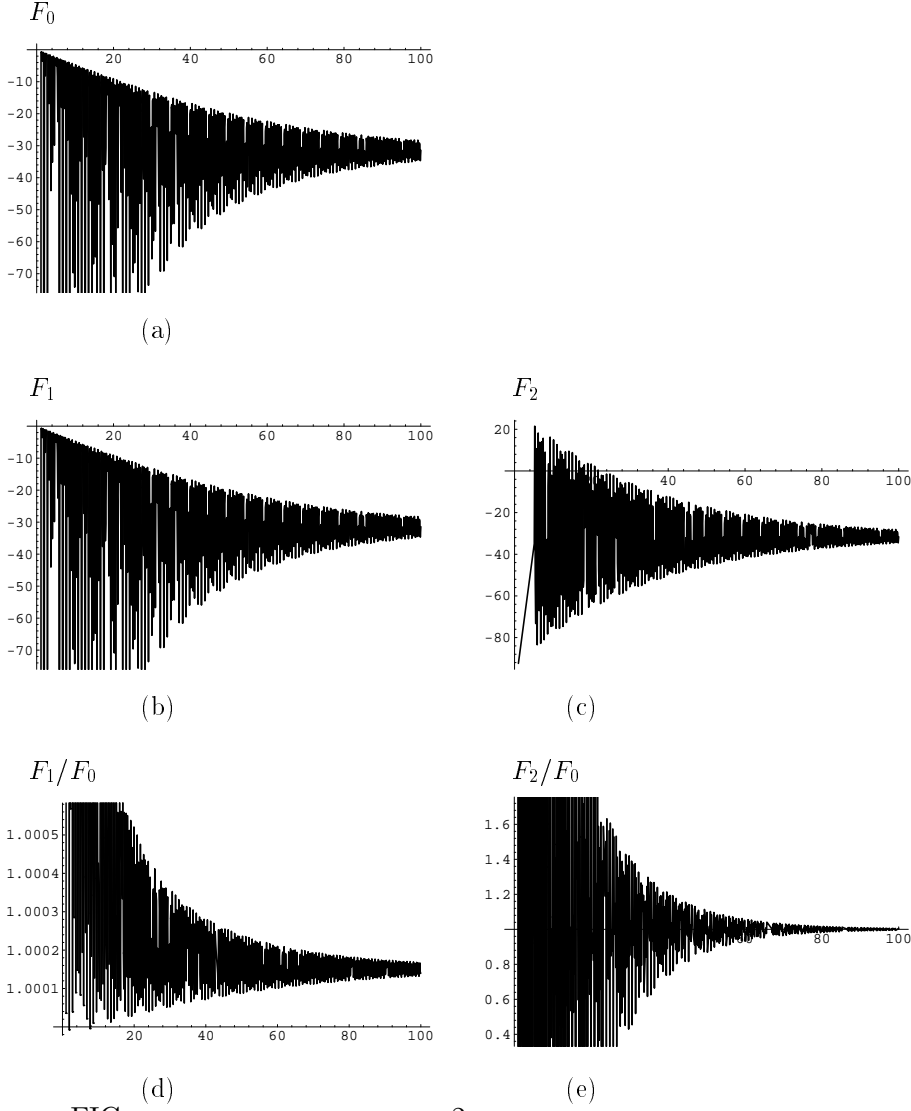}}
\caption{Polarisation functions in 2d QCD.
Here:
(a) $F_0={\rm Im}\frac{1}{\pi^2\mu^2}
\psi (-(q^2+\frac{Bq^2}{N_c\pi}
{\rm Log} (q^2/\bar q^2)+i Bq^2/N_c)/(\mu^2\pi^2 ))$,
(b) $F_1={\rm Im}\frac{1}{\pi\mu^2}
{\rm ctg}\pi(q^2+\frac{Bq^2}{N_c\pi}
\log {q^2/\bar q^2 }+iBq^2/N_c)/(\pi^2\mu^2)))$,
(c) $F_2=-\frac{1}{\pi\mu^2}
(1+2\exp{(-2Bq^2/\pi N_c\mu^2 )}\cos{(2q^2/(\pi\mu^2))})$,
and (d) for $F_1/F_0$, (e) for $F_2/F_0$.
We take $N_c=10$, $B=4.4\times 10^{-3}$ and $\mu=0.01$.
We use here obvious units of dimension.
}
\end{figure}
\newpage

\begin{figure}[htb]
{\epsfxsize3.0in\epsfbox{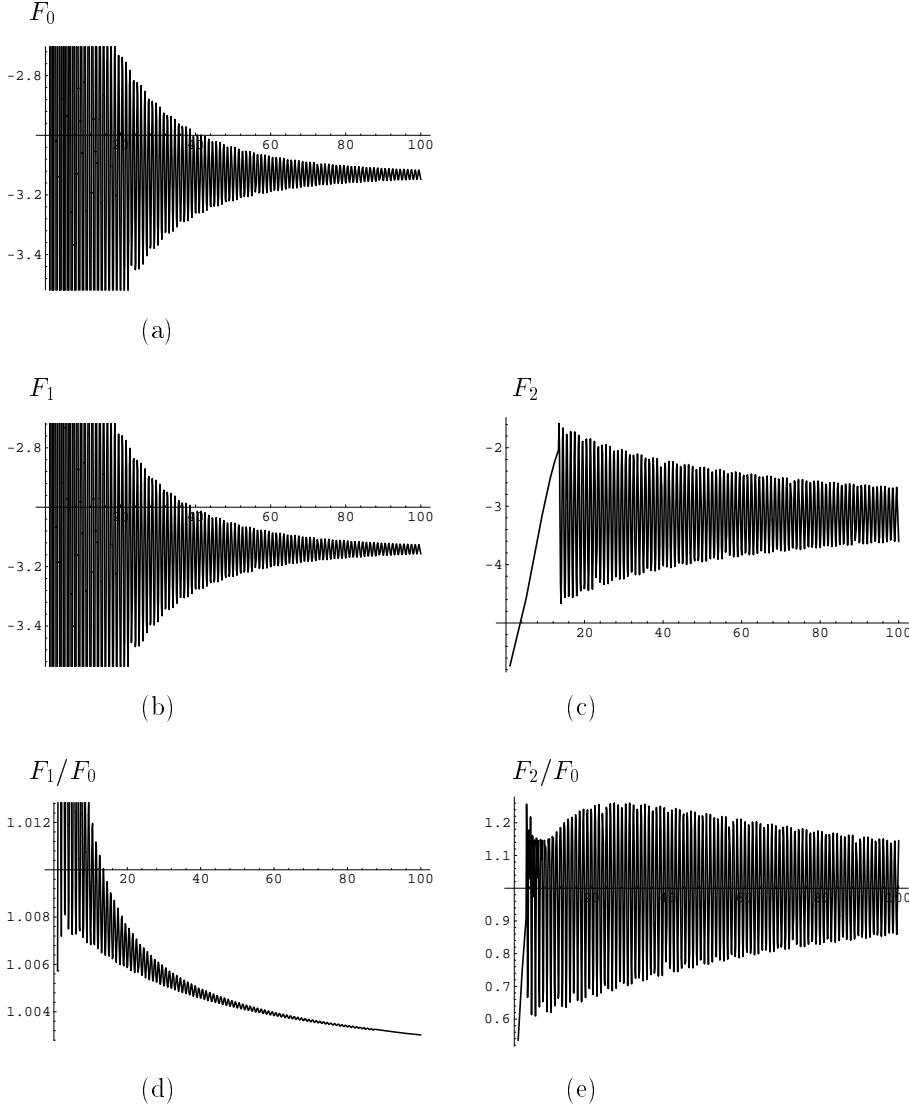}}
\caption{Polarisation functions in 4d QCD for $\gamma=-0.5$ and $A/N_c=0.3$.
(a) $F_0={\rm Im} \psi(-(q^2+ A/(N_c\pi) {q^2}^{(1+\gamma)} ctg(\pi \gamma)
+i A/(N_c\pi) {q^2}^{(1+\gamma)})))$,
(b) $F1=
\pi {\rm Im}ctg(\pi (q^2+A/(N_c\pi)  {q^2}^{(1+\gamma)} ctg(\pi \gamma)
+i  A/(N_c\pi){q^2}^{(1+\gamma)})))$,
(c) $F2=
-\pi (1+{\rm exp}(-2 A/(N_c\pi) {q^2}^{(1+\gamma)})
cos(2\pi (q^2+A/(N_c\pi)  {q^2}^{(1+\gamma)} ctg(\pi \gamma))))$,
and (d) for $F_1/F_0$, (e) for $F_2/F_0$. We use here obvious units 
of dimension.}
\end{figure}

\end{document}